\documentclass[aps,pre,floats,floatfix,twocolumn,showpacs,superscriptaddress]{revtex4}
\usepackage{latexsym,amsmath}
\usepackage{amsbsy}
\usepackage[pdftex]{graphicx}
\usepackage{amssymb}
\usepackage{epstopdf}
\usepackage[usenames]{color}

\begin{document}

\title{Clustering of random scale-free networks}

\author{Pol Colomer-de-Simon}
\affiliation{Departament de F{\'\i}sica Fonamental, Universitat de Barcelona, Mart\'{\i} i Franqu\`es 1, 08028 Barcelona, Spain}

\author{Mari{\'a}n Bogu{\~n}{\'a}}
\affiliation{Departament de F{\'\i}sica Fonamental, Universitat de Barcelona, Mart\'{\i} i Franqu\`es 1, 08028 Barcelona, Spain}

\date{\today}

\begin{abstract}

We derive the finite size dependence of the clustering coefficient of scale-free random graphs generated by the configuration model with degree distribution exponent $2<\gamma<3$. Degree heterogeneity increases the presence of triangles in the network up to levels that compare to those found in many real networks even for extremely large nets. We also find that for values of $\gamma \approx 2$, clustering is virtually size independent and, at the same time, becomes a {\it de facto} non self-averaging topological property. This implies that a single instance network is not representative of the ensemble even for very large network sizes.

\end{abstract}


\maketitle

Null models are critical to gauge the effect that randomness may have on the properties of systems in the presence of noise. It is therefore important to have the maximum understanding of the null model at hand, something not always easy to achieve. This is the case of the most used null model of random graphs, the configuration model (CM)~\cite{Bekessy:1972bp,Bender:1978xk,Molloy:1995rh,Molloy:1998od} 

Given a real network, the configuration model preserves the degree distribution of the real network, $P(k)$, whereas connections among nodes are realized in the most random way, always preserving the degree sequence --either the real one or drawn from the distribution $P(k)$-- and avoiding multiple and self-connections. In principle, the CM generates graphs without any type of correlations among nodes. For this reason, it is widely used in network theory to determine whether the observed topological properties of the real network might be considered as the product of some non trivial principle shaping the evolution of the system.

This program is severely hindered when the network contains nodes with degrees above the structural cut-off $k_s=\sqrt{\langle k \rangle N}$~\cite{Boguna:2004eh}, where $\langle k \rangle$ is the average degree and $N$ the size of the network. This is the case of scale-free networks with $P(k)\sim k^{-\gamma}$, $\gamma<3$, and a natural cut-off $k_c \sim N^{1/(\gamma-1)}$ most often found in real complex networks~\cite{NewmanBook:2010}. This apparently simple null model develops all sort of anomalous behaviors in this case, e. g., the appearance of strong non-trivial degree correlations among nodes~\cite{Burda:2003eo,Park:2003qy,Boguna:2004eh,Catanzaro:2005su}, difficulties in the sampling of the configuration space~\cite{Klein:2012}, or the presence of phase transitions between graphical and non-graphical phases~\cite{DelGenio:2011}, to name just a few.    

Clustering --or the presence of triangles in the network-- is yet another example of anomalous behavior associated to the CM. The importance of clustering as a topological property is related to the fact that nearly all known real complex networks have a very large number of triangles whereas the CM has a vanishingly small number in the thermodynamic limit. Of course, the absence of triangles is convenient from a theoretical point of view as it allows us to use generating functions techniques to solve many interesting problems~\cite{NewmanBook:2010}. However, given the empirical observations, it seems to be a quite unrealistic assumption. This has led to the common understanding that clustering observed in real networks cannot be explained by the CM and, thus, is the product of some underlying principle. While we fully agree with this statement, in this paper, we show that it must be taken with care. Indeed, depending on the heterogeneity of $P(k)$, the CM can generate, on average, nearly size-independent levels of clustering. Besides, in such cases, sample-to-sample fluctuations do not vanish when $N\rightarrow \infty$, meaning that the same degree sequence may generate either very high or very low levels of clustering, independently of the network size.

Clustering can be quantified using different metrics~\cite{Serrano:2006qj}. Here, we use the average clustering coefficient $C$, defined as the average (over nodes of degree $k \ge 2$) of the local clustering coefficient of single nodes $c_i=2 T_i/k_i (k_i-1)$, with $T_i$ the number of triangles attached to node $i$. In the absence of high degree nodes, the clustering coefficient of a random graph generated by the CM is given by~\cite{Newman:2003ka}
\begin{equation} 
C=\frac{\langle k(k-1) \rangle^2}{N \langle k \rangle^3},
\label{clustering1}
\end{equation}
and, therefore, vanishes very fast in the large system size. This is the reason why the tree-like character of networks generated by the CM has always been taken for granted. However, Eq.~(\ref{clustering1}) is clearly incorrect when the degree distribution is scale-free, as it predicts a behavior $C\sim N^{(7-3\gamma)/(\gamma-1)}$ that diverges for $\gamma<7/3$. Equation~(\ref{clustering1}) fails in this case because its derivation does not account for the structural correlations among degrees of connected nodes.
In this paper, we derive the correct scaling behavior of the clustering coefficient for scale-free random graphs with $2<\gamma<3$. 

The CM, as originally defined, defines a micro-canonical ensemble, in the sense that the degree of every single node is given a priori and, once the degree sequence is fully known, the network is assembled in the most random way while preserving the degree sequence. However, in the case of scale-free networks, this approach resists any analytic treatment. Instead, here we adopt a different strategy and work with the canonical ensemble of the CM. In this ensemble, each node is given not its actual degree but its expected degree. This relaxes the topological conditions to close the network and opens the door to an analytic treatment. Specifically, the model is defined as follows 
\begin{enumerate}
\item
Each node is assigned a hidden variable $\kappa$ drawn from the probability density $\rho(\kappa)=\propto \kappa^{-\gamma}$ with $1 \le \kappa \le \kappa_c$. The cut-off value $\kappa_c$ is, in principle, arbitrary. However, often $\kappa_c$ is the so-called natural cut-off, defined as the expected maximum value out of a sample of $N$ random deviates given from the probability density $\rho(\kappa)$. In the case of interest of a scale-free distribution, the natural cut-off scales as $\kappa_c \sim N^{1/(\gamma-1)}$.
\item
Each pair of nodes is visited once and connected with probability 
\begin{equation}
r\left(\frac{\kappa \kappa'}{\kappa_s^2}\right)= \frac{\kappa \kappa'}{\kappa_s^2} (1+\frac{\kappa \kappa'}{\kappa_s^2})^{-1},
\end{equation}
where $\kappa$ and $\kappa'$ are the hidden variables associated to each node, $\kappa_s=\sqrt{\frac{(\gamma-1)N}{(\gamma-2)\bar{k}_{min}}}$, and $\bar{k}_{min}$ is the expected minimum degree of the network. The particular form chosen for the connection probability ensures that the entropy of the ensemble is maximal~\cite{Garlaschelli:2008fd,Anand:2009am,Anand:2011}.
\end{enumerate}
It can be shown that the average degree of a node with hidden variable $\kappa$ is $\bar{k}(\kappa) \propto \kappa$~\cite{Park:2003qy,Boguna:2003um,serrano:2011}. Thus, we can think of $\kappa$ and $\rho(\kappa)$ as the degree and degree distribution, respectively. 

Parameter $\kappa_s$ is a structural cut-off defining the onset of structural correlations, that is, nodes with expected degrees below $\kappa_s$ are connected with probability $r(\frac{\kappa \kappa'}{\kappa_s^2}) \approx \frac{\kappa \kappa'}{\kappa_s^2}$ and, therefore, are uncorrelated at the level of degrees. As a consequence, the global level of correlations present in the system is controlled by the cut-off $\kappa_c$. Whenever $\kappa_c < \kappa_s$ the resulting network is fully uncorrelated whereas for $\kappa_c \ge \kappa_s$ correlations are necessary to close it. In this paper, we are interested in the range $\kappa_s \le \kappa_c \le N^{1/(\gamma-1)}$.

Using the formalism developed in\cite{Boguna:2003um}, the local clustering coefficient of a node with hidden variable $\kappa$ can be written as
\begin{equation}
c(\kappa)=\displaystyle{
\frac{
\displaystyle{
\int_{\frac{1}{\kappa_s}}^{\frac{\kappa_c}{\kappa_s}}}
\int_{\frac{1}{\kappa_s}}^{\frac{\kappa_c}{\kappa_s}} 
\frac{1}{(x y)^{\gamma}} r\left(\frac{\kappa x}{\kappa_s}\right) r\left(x y \right) r\left(\frac{\kappa y}{\kappa_s}\right)dx dy}{\left[
\displaystyle{
\int_{\frac{1}{\kappa_s}}^{\frac{\kappa_c}{\kappa_s}}}
x^{-\gamma}r\left(\frac{\kappa x}{\kappa_s}\right)dx \right]^2}.
}
\label{eq:clustering}
\end{equation}
The average clustering coefficient is computed from $c(\kappa)$ as $C=\int \rho(\kappa) c(\kappa) d \kappa$. However, since $c(\kappa)$ is a bounded monotonously decreasing function its major contribution to $C$ comes from nodes with small degree, i. e., low $\kappa$. Therefore, to find the correct scaling behavior it suffices to evaluate $c(\kappa)$ in the domain $\kappa<<\kappa_s$. In this case, the maximum value within the domain of integration $[1/\kappa_s,\kappa_c/\kappa_s]$ of the arguments $\kappa x/\kappa_s$ and $\kappa x/\kappa_s$ in Eq.~(\ref{eq:clustering}) is of order $\mathcal{O}(\kappa_c/\kappa_s^2)$, which goes to zero in the thermodynamic limit. We can, thus, approach $c(\kappa)$ as
\begin{equation}
c(\kappa)\approx \frac{(\gamma-2)^2}{\kappa_s^{2(\gamma-2)}(1-\kappa_c^{2-\gamma})^2} 
\int_{\frac{1}{\kappa_s}}^{\frac{\kappa_c}{\kappa_s}}
\int_{\frac{1}{\kappa_s}}^{\frac{\kappa_c}{\kappa_s}}
\frac{(xy)^{2-\gamma}}{1+xy}dx dy,
\end{equation}
which becomes independent of $\kappa$. After some manipulation, this expression becomes
\begin{widetext}
\begin{equation}
c(\kappa) \approx \frac{(\gamma-2)^2}{\kappa_s^{2(\gamma-2)}(1-\kappa_c^{2-\gamma})^2}
\left[ 2 \psi(\gamma) \ln{\left(\frac{\kappa_c}{\kappa_s} \right)} 
+\theta(\gamma)
+\left(\frac{\kappa_s}{\kappa_c}\right)^{2(\gamma-2)} \Phi\left( -\left(\frac{\kappa_s}{\kappa_c}\right)^2,2,\gamma-2 \right)
\right.
\label{clustering_main}
\end{equation}
\[
\left.
-2 \left(\frac{\kappa_c}{\kappa_s^2}\right)^{3-\gamma} \Phi\left(-\frac{\kappa_c}{\kappa_s^2},2,3-\gamma \right)
+ \frac{1}{\kappa_s^{6-2\gamma}} \Phi\left(-\frac{1}{\kappa_s^2},2,3-\gamma \right)
 \right]
\]
\end{widetext}
where
\[
\psi(\gamma)=\Phi(-1,1,3-\gamma)+\Phi(-1,1,\gamma-2),
\]
\[
\theta(\gamma)=-\pi^2 \cot{\pi \gamma} \csc{\pi \gamma},
\]
and $\Phi(z,a,b)$ is the transcendent Lerch function~\cite{Gradstein:2000zp}. This expression, although involved at first glance, it is convenient because in the range $\kappa_s\le \kappa_c \ll \kappa_s^2$ the arguments of the three transcendent Lerch functions in it go to $0^-$ in the limit $\kappa_s \rightarrow \infty$, in which case we know that $\Phi(-z^2,a,b)\sim b^{-a}$ for $z\rightarrow 0$. We then find the asymptotic behavior
\begin{equation}
c(\kappa) \sim 
\frac{(\gamma-2)^2}{\kappa_s^{2(\gamma-2)}}  
\left\{
\begin{array}{lr}
\theta(\gamma)+\Phi(-1,2,\gamma-2) & \kappa_c=\kappa_s\gg1\\[0.5cm]
2  \psi(\gamma) \ln{\left(\frac{\kappa_c}{\kappa_s} \right)} & \kappa_c\gg \kappa_s\gg 1.
\end{array}
\right.
\end{equation}
The first line in this equation recovers the result found in~\cite{Catanzaro:2005su} for scale-free networks without structural correlations --$c(\kappa) \sim N^{2-\gamma}$ when $\kappa_s \sim N^{1/2}$-- whereas the second line predicts $c(\kappa) \sim N^{2-\gamma}\ln{N}$ when $\kappa_c \sim N^{1/(\gamma-1)}$, which corrects the incorrect scaling behavior predicted by Eq.~(\ref{clustering1}) in this case. 
\begin{figure}[t]
\centerline{\includegraphics[width=3.5in]{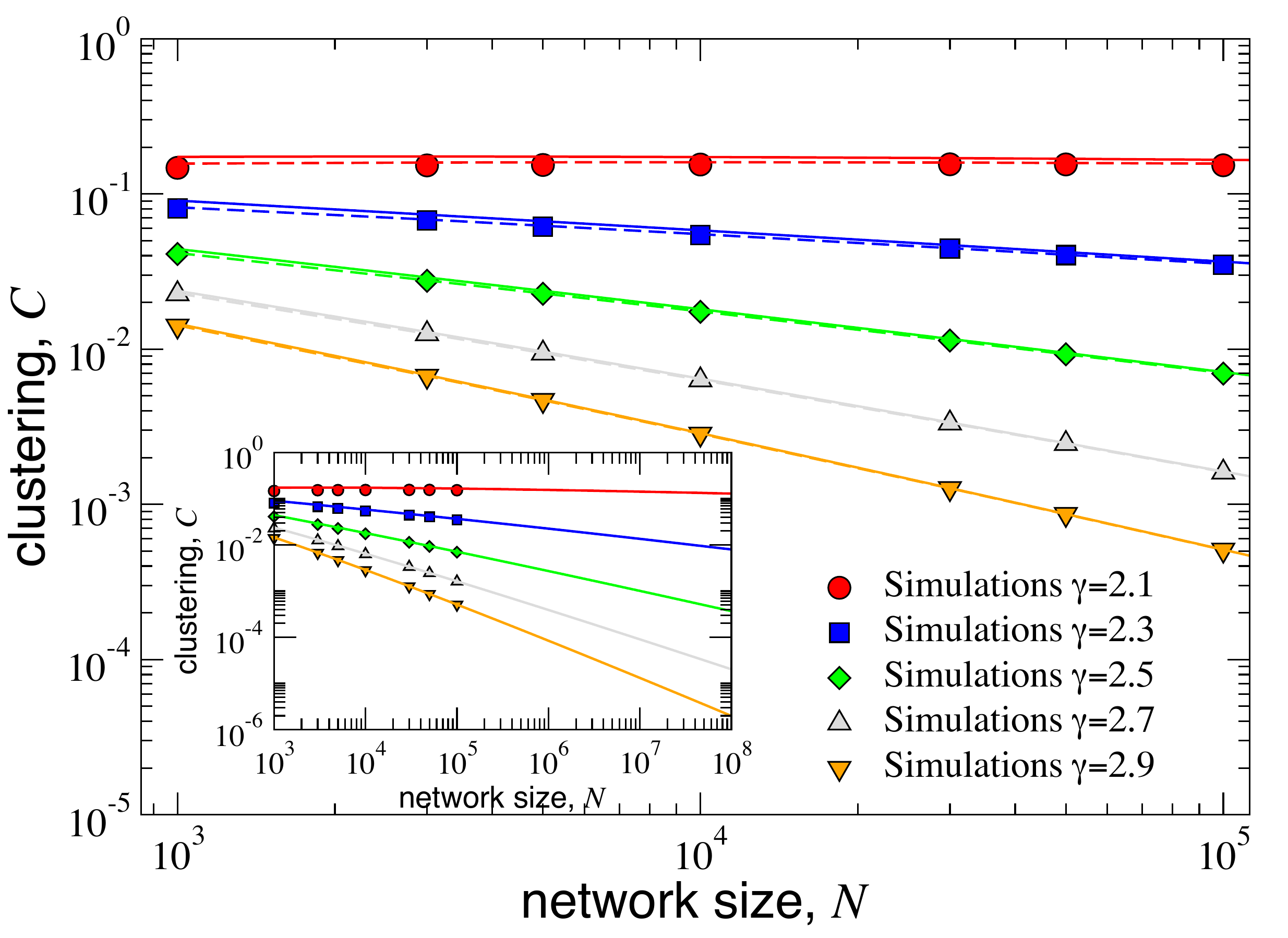}}
\caption{Clustering coefficient as measured in numerical simulations for different values of $\gamma$ and size $N$ with $\bar{k}_{min}=2$ and $\kappa_c=N^{1/(\gamma-1)}$. Each point is an average over $10^4$ different network realizations. Dashed lines are the numerical solution of Eq.~(\ref{eq:clustering}) and solid lines are the approximate solution given by Eq.~(\ref{clustering_main}). The inset shows an extrapolation up to size $N=10^8$ using Eq.~(\ref{clustering_main}).}
\label{fig1}
\end{figure}
\begin{figure}[t]
\centerline{\includegraphics[width=3.5in]{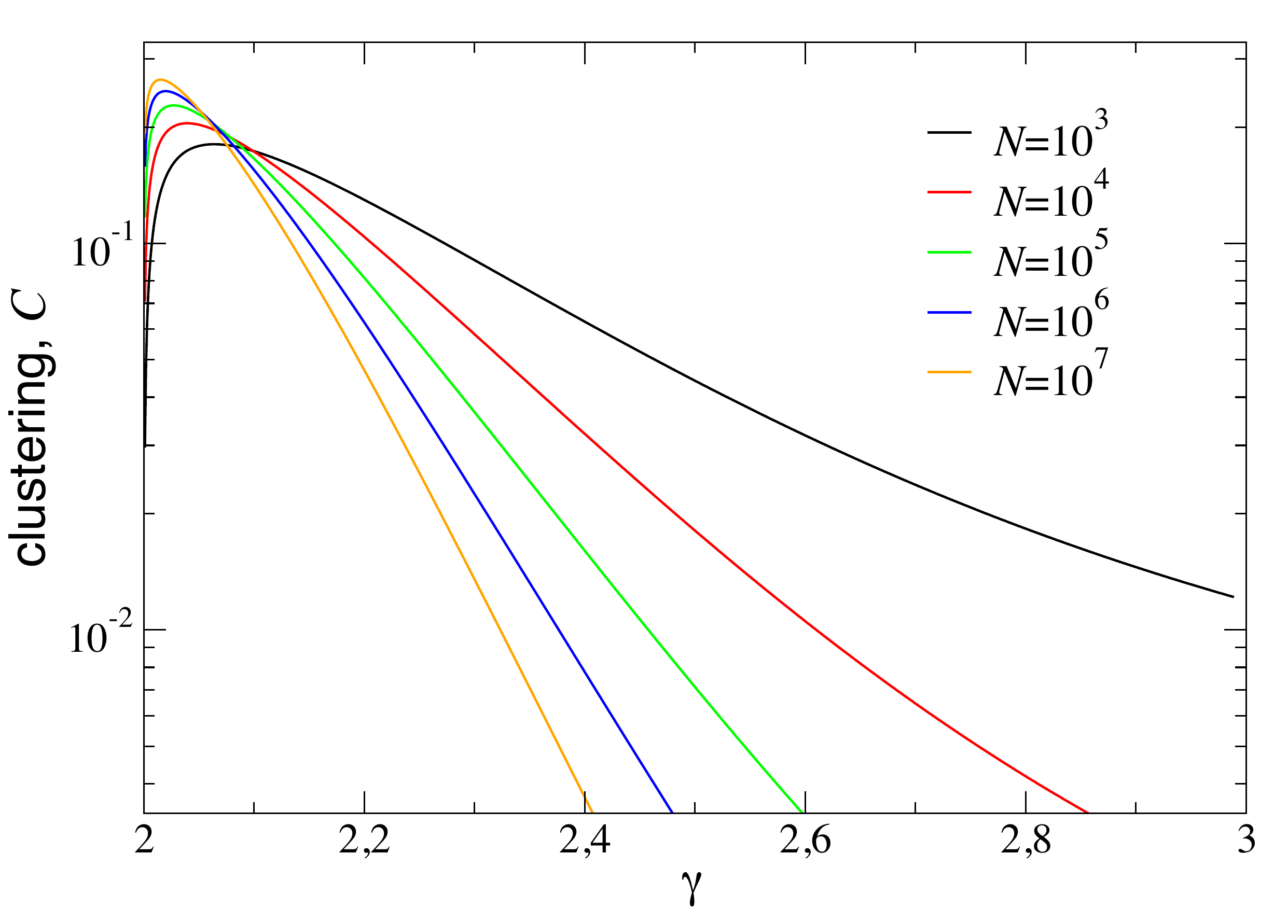}}
\caption{Clustering coefficient as a function of $\gamma$ for different network sizes. Curves are evaluated from Eq.~(\ref{clustering_main}) with $\bar{k}_{min}=2$ and $\kappa_c=N^{1/(\gamma-1)}$.}
\label{fig2}
\end{figure}

\begin{figure}[t]
\centerline{\includegraphics[width=3.5in]{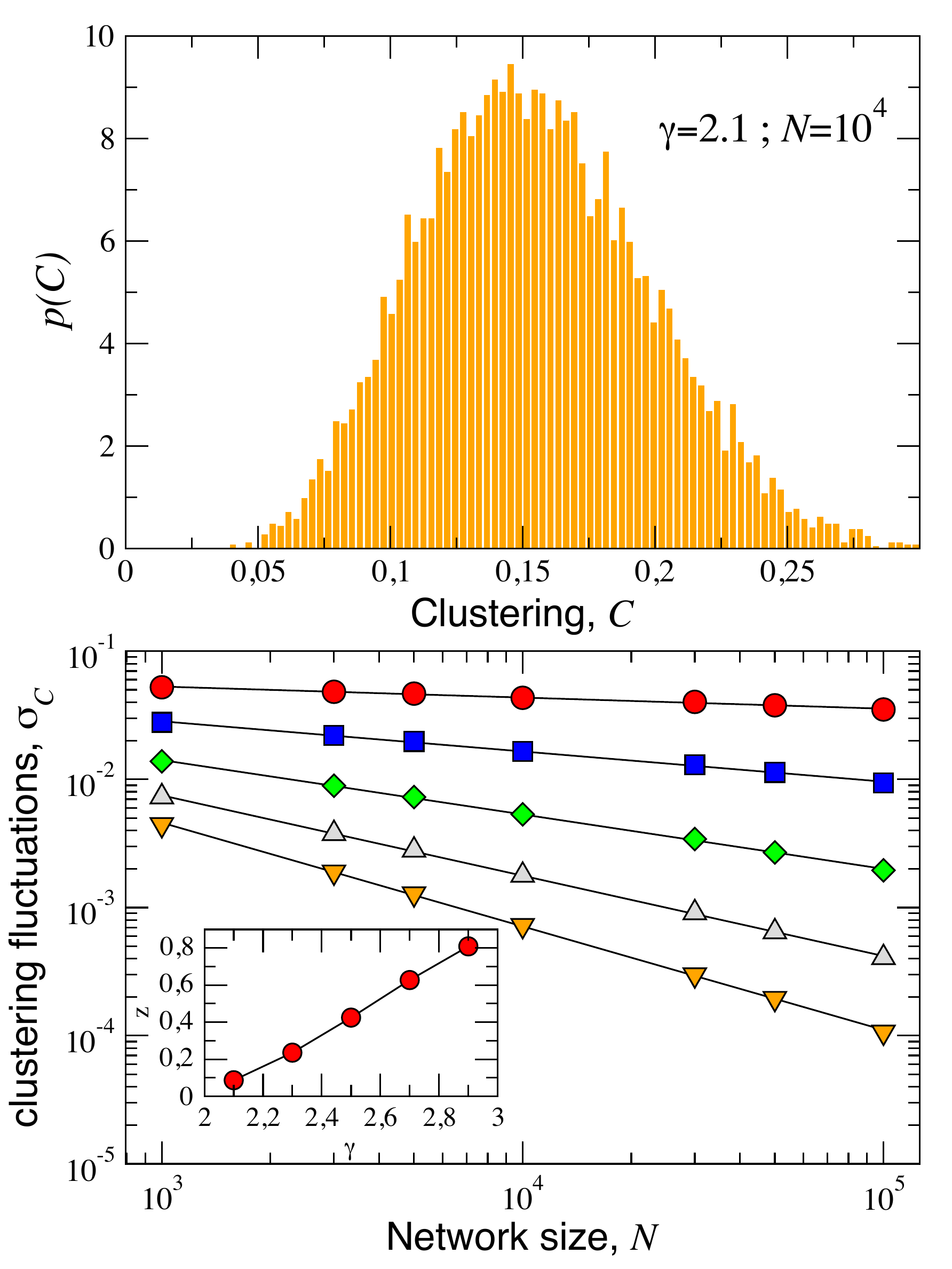}}
\caption{Sample to sample fluctuations. The top plot shows the probability density function of the clustering coefficient as obtained from $10^4$ network realizations for $\bar{k}_{min}=2$, $\kappa_c=N^{1/(\gamma-1)}$, $\gamma=2.1$, and $N=10^4$. The bottom plot shows the standard deviation of this pdf for different values of $\gamma$ as a function of the network size. Solid lines are power law fits of the form $\sigma_C \sim N^{-z}$. The exponent $z$ is shown in the inset.}
\label{fig3}
\end{figure}

Figure~\ref{fig1} shows a comparison between numerical simulations, the numerical solution of Eq.~(\ref{eq:clustering}), and the approximate solution given by Eq.(\ref{clustering_main}), showing a very nice agreement. Interestingly, for $\gamma=2.1$, clustering remains nearly constant in the range of sizes $10^3-10^5$ and even increases slightly for small sizes. This is a consequence of the slow decay of the term $\kappa_s^{2(2-\gamma)}$ combined with the diverging logarithmic term in the numerator and functions $\psi(\gamma)$ and $\theta(\gamma)$, which diverges in the limit $\gamma \rightarrow 2$. In the inset of Fig.~\ref{fig1}, we show the extrapolation of the clustering coefficient for sizes up to $10^8$ evaluated with Eq.~(\ref{clustering_main}). In the case of $\gamma=2.1$, this figure makes evident the extremely slow decay --nearly absent-- with the system size. This implies that, in practice, clustering cannot be removed from the network even in very large networks when $\gamma \approx 2$. It is, thus, not clear whether the tree-like approximation, customarily used to solve problems on random graphs, can be applied in this case. In this situation, one should use alternative approaches, like the one developed in~\cite{serrano:2011}. These results are particularly relevant due to the abundance of real networks with values of $\gamma \approx 2$. 

It is also interesting to study the behavior of clustering as a function of $\gamma$ for a fixed network size. Figure~\ref{fig2} shows this behavior for different values of $N$. For each size, there is an optimal value of $\gamma$ where clustering is maximal. In the case of $\bar{k}_{min}=2$ and $N\ge 10^3$, there is critical value $\gamma_{crit} \approx 2.1$ below which clustering increases with the network size up to a maximum and then slowly decreases. 

Up to this point, we have been concerned only with the ensemble average of the clustering coefficient. However, the CM ensemble shows strong sample-to-sample fluctuations. Figure~\ref{fig3} shows the probability density function of the clustering coefficient obtained out of a sample of $10^4$ different networks generated by the canonical version of the CM. As it can be observed, clustering may take values in the range $[0.05,0.25]$ quite easily. Figure~\ref{fig3} also shows the standard deviation $\sigma_C$ as a function of network size and for different values of $\gamma$. In all cases, fluctuations decay as a power law of the system size, $\sigma_C \sim N^{-z}$, with an exponent $z<1$. Interestingly, for $\gamma=2.1$, the exponent $z$ takes a very small value ($z \approx 0.1$) that, when combined with the behavior of $C$ as a function of $N$ results in a coefficient of variation nearly constant. This implies that, in this range of values of $\gamma$, clustering is {\it de facto} a size-independent but non self-averaging property. That is, a single network instance is not a good representative of the ensemble even for very large network sizes.

The presence of triangles in real networks play an important role in many processes taking place on top of them, e. g. , percolation phenomena, epidemic spreading, synchronization, etc. It is, therefore, important to have full control over the most simple network ensembles that are used as null models to assess the presence of underlying principles shaping the topology of the system. In this paper, we have found the correct scaling behavior of the clustering coefficient of the ensemble of scale-free random graphs with $2<\gamma<3$. Interestingly, for values of the exponent $\gamma \approx 2$, clustering remains nearly constant up to extremely large network sizes. However, in this case, clustering is not self-averaging. This means that when comparing real networks against the CM, it is not enough to generate a single instance network, as it may result in either a very low or high level of clustering even for very large network sizes. These results are particularly important as the exponent value $\gamma \approx 2$ seems to be --for yet unknown reasons-- the rule rather than the exception in real systems.

\section*{Acknowledgements}

We thank M. \'Angeles Serrano for useful comments and suggestions. This work was supported by MICINN Project No.\ FIS2010-21781-C02-02; Generalitat de Catalunya grant Nos.\ 2009SGR838; ICREA  Academia prize 2010, funded by the Generalitat de Catalunya.


\end{document}